\documentclass[review,10pt]{elsarticle}
\usepackage{lineno,hyperref}
\modulolinenumbers[5]
\usepackage{amsmath}
\usepackage{amsfonts}
\usepackage{amssymb}
\usepackage{lmodern}
\usepackage{xfrac}
\usepackage{graphicx}
\usepackage{subfig}
\usepackage{epsfig}
\usepackage{epstopdf}
\usepackage{textcomp}
\usepackage{multirow}
\usepackage{setspace}
\usepackage{fullpage}
\usepackage{float}
\usepackage{caption}
\usepackage{siunitx}
\usepackage{xcolor}
\usepackage{tabularx, booktabs}

\biboptions{sort&compress}

\begin{document}

\begin{frontmatter}
\title{Exploring large language models for microstructure evolution in materials} 

\cortext[contrib]{These authors contributed equally.}
\cortext[mycorrespondingauthor]{Corresponding author.}
\author{Prathamesh Satpute\corref{contrib}}
\author{Saurabh Tiwari\corref{contrib}}
\author{Maneet Gupta}
\author{Supriyo Ghosh\corref{mycorrespondingauthor}}
\ead{supriyo.ghosh@mt.iitr.ac.in; gsupriyo2004@gmail.com}
\address{Department of Metallurgical and Materials Engineering, Indian Institute of Technology, Roorkee, UK 247667, India}



%


\begin{abstract}
There is a significant potential for coding skills to transition fully to natural language in the future. In this context, large language models (LLMs) have shown impressive natural language processing abilities to generate sophisticated computer code for research tasks in various domains. We report the first study on the applicability of LLMs to perform computer experiments on microstructure pattern formation in model materials. In particular, we exploit LLM's ability to generate code for solving various types of phase-field-based partial differential equations (PDEs) that integrate additional physics to model material microstructures. The results indicate that LLMs have a remarkable capacity to generate multi-physics code and can effectively deal with materials microstructure problems up to a certain complexity. However, for complex multi-physics coupled PDEs for which a detailed understanding of the problem is required, LLMs fail to perform the task efficiently, since much more detailed instructions with many iterations of the same query are required to generate the desired output. Nonetheless, at their current stage of development and potential future advancements, LLMs offer a promising outlook for accelerating materials education and research by supporting beginners and experts in their physics-based methodology. We hope this paper will spur further interest to leverage LLMs as a supporting tool in the integrated computational materials engineering (ICME) approach to materials modeling and design.
\end{abstract}

\end{frontmatter}

\section{Introduction}
Large language models (LLMs) are a conversational, generative artificial intelligence (AI) tool that offers massive potential for transforming science and society by redefining human-computer interactions~\cite{van2023chatgpt,schulze2024empirical}. In principle, LLMs take an input text prompt through a free chat interface and spontaneously, as if by magic, write a human-like response virtually on \textit{any} topic. Some of the well-known LLMs are given in Table~\ref{table:llm}. These models are based on massive machine learning (ML) architectures (specifically transformer neural networks~\cite{geron2022,vaswani2017attention}) pre-trained on a massive corpus of text data from diverse internet sources like Common Crawl~\cite{commoncrawl} and Wikipedia~\cite{wikipedia}, each contributing to varying degrees of weightage to different aspects of the model's knowledge~\cite{alto2023modern}. This allows LLMs to perform natural language processing (NLP) tasks, including human-like conversation and contextual understanding. The LLMs continue to learn in a self-supervised manner and correct errors by themselves through the process of reinforcement learning from human feedback (RLHF)~\cite{lei2024materials,liu2023generative}. Therefore, for a given task, one needs to craft the question prompts more carefully, allowing LLMs to be more effective at handling complex tasks and generating better and more reliable answers. It is even possible to obtain the desired output with ``zero-shot'' prompting (i.e, giving no example of how the task should be solved) or typically with a series of prompts or ``chain prompting''. In addition, prompt processing, such as reducing the length of prompts, can result in faster inference times and lower computational costs~\cite{llmlingua}. Thus, LLMs are increasingly becoming useful to complete research tasks quickly that include generating data, code, models, essays, and literature reviews. However, the performance of LLMs is still not well-tested for specialized research purposes that require domain-specific knowledge. This work explores how far we can push LLMs to generate desired output for complex research tasks in an example materials science domain.  

\begin{table}[htbp]
\centering
\caption{Overview of major LLMs. Model size indicates LLM's complexity and capacity for learning. The token limit refers to the maximum length of input/output text it can handle in a single prompt.}\label{table:llm}
\begin{tabular}{|l|l|l|l|l|}
\hline
\textbf{Model Name} & \textbf{Developer} & \textbf{Release Date} & \textbf{Model Size (Parameters)} & \textbf{Token limit} \\ \hline
ChatGPT-3.5~\cite{gpt3} & OpenAI & June 2020 & 175 billion & 4096 \\ 
ChatGPT-4 (Turbo)~\cite{gpt4} & OpenAI & March 2023 & 1 trillion & 128000 \\ 
Grok~\cite{grok} & xAI & November 2023 & 314 billion & 8000 \\ 
Gemini~\cite{gemini} & Google AI & March 2023 & 2 billion - 7 billion & 32000 \\ 
Llama-2~\cite{llama} & Meta & July 2023 & 7 - 70 billion & 2048 \\
Llama-3~\cite{llama} & Meta & April 2024 & 8 - 70 billion & 8000 \\
Copilot~\cite{copilot} & Microsoft & February 2023 & Based on GPT-3.5 and 4 & 4096\\
Cluade-3 Sonnet~\cite{cluade} & Anthropic & February 2024 &  $>$ 20-70 billion & 200000   \\
Mistral~\cite{mistral} & Mistral AI & September 2023 & 7-12 billion & 32000   \\ \hline 
\end{tabular}
\end{table}

As computational scientists, we are always eager to embrace new AI tools that will save us time and offer a great deal of convenience. Among the LLMs (Table~\ref{table:llm}), ChatGPT (Chat Generative Pre-trained Transformer) has garnered significant attention from the public, academia, industry, and media. As we will show later, ChatGPT outperforms other LLMs when it comes to coding and data analysis. With over 175 billion ML parameters, its ability to understand/generate text, analyze data, generating code, and even run Python codes sets it apart from other LLMs. The ChatGPT-4, the latest iteration of the GPT model (as of April 1, 2024), is even more powerful as it is pre-trained with 1 trillion data set with increased token/context capacity, dramatically enhancing its performance in solving complex tasks in applications reaching diverse fields including mathematics~\cite{frieder2024mathematical,koceska2023can,orlando2023assessing}, coding~\cite{liu2023generative,orlando2023assessing}, medicine~\cite{dave2023chatgpt}, biology~\cite{lubiana2023ten}, and even law~\cite{choi2021chatgpt}. Also, the incredible capability of ChatGPT-4 in working with multimodal data (i.e., images, text, audio, and video) and conversational interactivity offer a promising outlook for scientific computing. In this work, we use ChatGPT-4 as a representative LLM to explore its knowledge within the realm of computational materials science, emphasizing its potential benefits efficiently. In the context of materials science, only a few published studies (three as of April 1, 2024) demonstrated applications of ChatGPT, particularly in atomistic materials modeling for generating/querying different crystal symmetry structures useful for density functional theory calculations~\cite{hong2023chatgpt,liu2023generative,deb2024chatgpt}. However, in the integrated computational materials engineering (ICME) framework, materials models at different length scales are linked together that fully capture material behavior in technological applications. In this work, we examine several popular LLMs, including ChatGPT, to explore their understanding of computational materials science at the microscale, particularly the modeling of microstructure evolution described by complex partial differential equations (PDEs). There have been a few previous studies on the mathematical and coding capabilities of ChatGPT for solving differential equations and PDEs~\cite{koceska2023can,orlando2023assessing}. However, these general studies are not explicitly related to materials applications. To our knowledge, this is the first attempt to systematically explore ChatGPT as an alternative to microstructure evolution modeling.

On the micro-to-meso scale, a critical factor in guaranteeing the superior performance of materials is to engineer the complex phase organizations and interface structures in the microstructure patterns generated during phase transformations. The phase-field (PF) method~\cite{chen2002,moelans2008,boettinger2002,steinbach2009} is the most powerful technique for the study of the thermodynamics and kinetics of interface migration and the associated microstructural evolution. A particular strength of the method is that the sharp interface is modeled by a diffuse interface with a finite width, which reduces to a set of nonlinear PDEs that are handled by standard numerical techniques (e.g., finite difference discretization method with explicit time marching scheme). In principle, these PDEs are the equations of motion obtained by minimizing the material's free energy function with respect to the field variables representing physical or phenomenological microstructure descriptors. Depending on the physics being modeled, these variables can be conserved (e.g., concentration) and non-conserved (e.g., phase fraction) order parameters. In a microstructure, these fields vary smoothly within the diffuse interface between their (scalar) values in the adjacent bulk solid and liquid phases. This approach avoids explicit tracking of the interface, and thus, the complex evolution of the interface, as seen in complex geometries and topology changes in materials, can be modeled efficiently. The temporal evolution of the field variables obtained by solving the PDEs represents the microstructure patterns. In this work, we explore the process of ChatGPT for solving several well-know PDEs that are central to materials science. We mainly address the following key questions: 
\begin{itemize}
    \item Can the computer experiments on microstructure pattern formation be outsourced to LLMs?
    \item How do the LLMs impact traditional materials education and research?
\end{itemize}

The structure of the rest of the article is as follows. In Sec.~\ref{sec:examples}, we label four example problems that we test with ChatGPT. In Sec.~\ref{sec:results}, we present the outputs generated by ChatGPT for all the case studies and compare their accuracy against our simulations using in-house codes. In Sec.~\ref{sec:summary}, we summarize and conclude with a detailed outlook on the present topic.

\section{General Approach}\label{sec:examples}
\begin{figure}
    \centering
   \includegraphics[width=\textwidth]{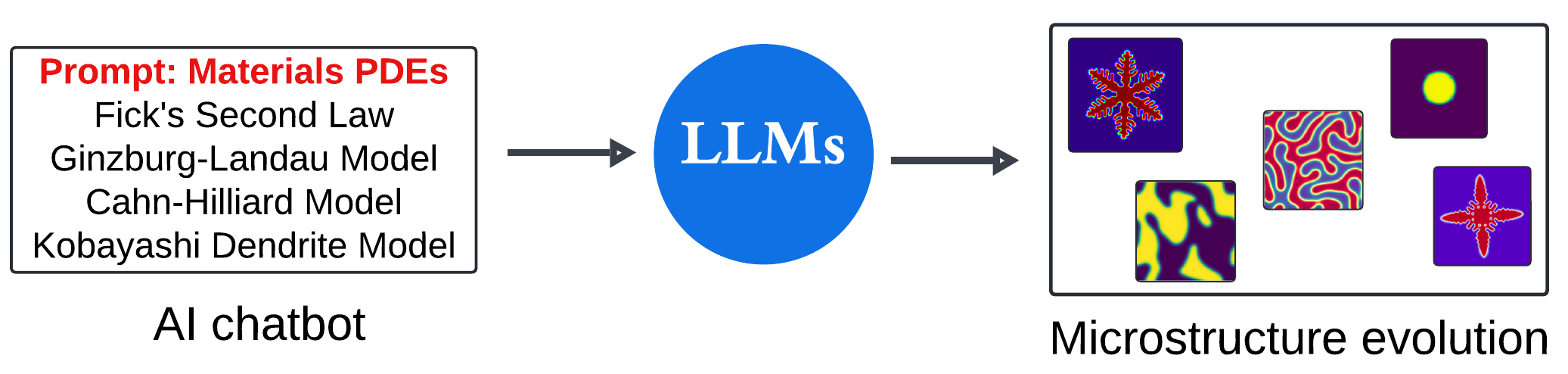}
    \caption{The schematic summarizes our workflow. The artificial intelligence (AI) chatbot provides a user interface to interact with large language models (LLMs) as a ``black box''. We input a partial differential equation (PDE) with appropriate initial and boundary conditions and material parameters as a prompt in the chatbot of a given LLM like ChatGPT (Table~\ref{table:llm}). Then, it generates a Python code to solve the given PDE and produces the output in terms of the time evolution of the microstructural pattern formation.}\label{fig:schematic}
\end{figure}

To examine the capability of LLMs, we select four benchmark phase-field model problems~\cite{pf_benchmark,pf_benchmark2} (or PDEs) with increasing order of model complexity in terms of additional physics. The first problem we study is the diffusion equation~\cite{callister} (i.e., Fick's second law); the second example is the Cahn-Hilliard equation~\cite{Cahn1958}; the third problem is the coupled Cahn-Hilliard and Allen-Cahn~\cite{Allen1979} (or Ginzburg-Landau) equations; the fourth problem is the dendritic crystal growth in pure systems by Kobayashi~\cite{kobayashi1993}. In this way, we focus on a single, fundamental aspect of physics (diffusion) in the first problem and increase the model complexity in terms of additional physics (diffusion and phase separation) in the second example, and so on. We also use simplified formulations of the above systems with dimensionless parameters to make the tests more straightforward to implement in ChatGPT. Also, we use one and two-dimensional simple domain geometries with suitable initial and boundary conditions so that the governing physics can be simulated without making the test problems unreasonably large or computationally demanding. In this way, we will be able to capture the essential physical behavior (e.g., solute diffusion, two-phase growth, coarsening, and solidification) in a vast majority of materials and models. A schematic of our workflow is given in Fig.~\ref{fig:schematic}. 

The LLMs, including ChatGPT, can be used as a mathematical tool for solving PDEs~\cite{frieder2024mathematical,koceska2023can,orlando2023assessing}. When we typically solve a PDE using analytical methods or computer experiments, the solution of the PDE is sought under certain initial and boundary conditions for the given material-specific parameters. Similarly, to solve a given PDE in ChatGPT, it needs to know the explicit form of the equation, the initial and boundary conditions, and the material parameters. Also, the user must specify the numerical implementation scheme the ChatGPT will employ. For example, we use an explicit Euler finite difference algorithm throughout this study. In a typical PDE solver, the above input parameters and constraints must be specified before the PDE can be solved. This is essentially what we input in the prompt to solve PDEs, as shown in Figs.~\ref{fig:ficks}-\ref{fig:dendrite}, providing the equations and parameters formatted in natural language. As a response, ChatGPT produces the output and visualization of the results.

 Besides the AI chatbot provided by several LLMs (Table~\ref{table:summary}), we also explore Hugging Face's~\cite{hugging} Hugging Chat interface to test out many excellent ready-to-use LLMs such as Llama-2~\cite{llama}, Llama-3~\cite{llama}, and Mistral~\cite{mistral} to solve material PDEs. In these LLMs, we use the default inference-time model parameters such as \texttt{Temperature} = 0.6, \texttt{Top-P} = 0.9, and \texttt{Top-K} = 50 as a reference. It is clear that ChatGPT-4 outperforms other popular LLMs when it comes to generating the correct complete code and the ability to execute it. Therefore, we instruct ChatGPT to do code writing to solve the given PDEs, noting that it can write and run Python codes. Also, for each task, we implement the identical problem on the client-side and compare the execution results with the ChatGPT outputs. To solve PDEs corresponding to each case study, we enter a text prompt in its AI chatbot, describing the PDEs and parameters to code in Python. Then, the ChatGPT responds with an explanation of the topic, algorithmic description, the Python program, and visualization/analysis of the execution results. To present these results in a concise manner, we here only report the prompt given and the output visualization files. We include the ChatGPT-generated codes for each task in the supplementary material.  

\begin{table*}[ht]
\centering
\caption{Summary of the capability of various LLMs (accessed as of March 31, 2024) on the four input prompts correspond to each task is presented with Y (Yes) and N (No). We access the latest Llama family model, Llama-3~\cite{llama}, from Hugging Face~\cite{hugging} on May 24, 2024.}\label{table:summary}
\begin{tabularx}{\textwidth}{|>{\raggedright\arraybackslash}X|c|>{\raggedright\arraybackslash}X|>{\raggedright\arraybackslash}X|}
\hline
\textbf{Large Language AI Model} & \textbf{Full Code (Python)} & \textbf{Correct Code (Task/Prompt 1-4)} & \textbf{Execution Ability/Data Visualization (Task/Prompt 1-4)} \\ 
\midrule
GPT-3.5~\cite{gpt3}  & Y                           & Y, N, N, N                         & N, N, N, N                                                   \\ 
GPT-4~\cite{gpt4}    & Y                           & Y, Y, Y, Y                         & Y, Y, Y, Y                                                   \\ 
Gemini~\cite{gemini}  & Y                           & Y, N, N, N                         & N, N, N, N                                                   \\ 
Llama-2~\cite{llama}   & Y                           & N, N, N, N                         & N, N, N, N                                                   \\ 
Llama-3~\cite{llama}   & Y                           & Y, N, N, N                         & N, N, N, N                                                   \\ 
Copilot~\cite{copilot}  & Y                           & Y, Y, N, N                         & N, N, N, N                                                   \\ 
Claude-3~\cite{cluade}   & Y                           & Y, Y, N, N                         & N, N, N, N                                                  \\ 
Mistral~\cite{mistral}   & Y                           & N, N, N, N                         & N, N, N, N                                                   \\ 
\bottomrule
\end{tabularx}
\end{table*}

\section{Simulation examples}\label{sec:results}
\subsection{Task 1: Transient solute diffusion equation}\label{sec:task1}
\begin{figure}[ht]
\includegraphics[width=\textwidth]{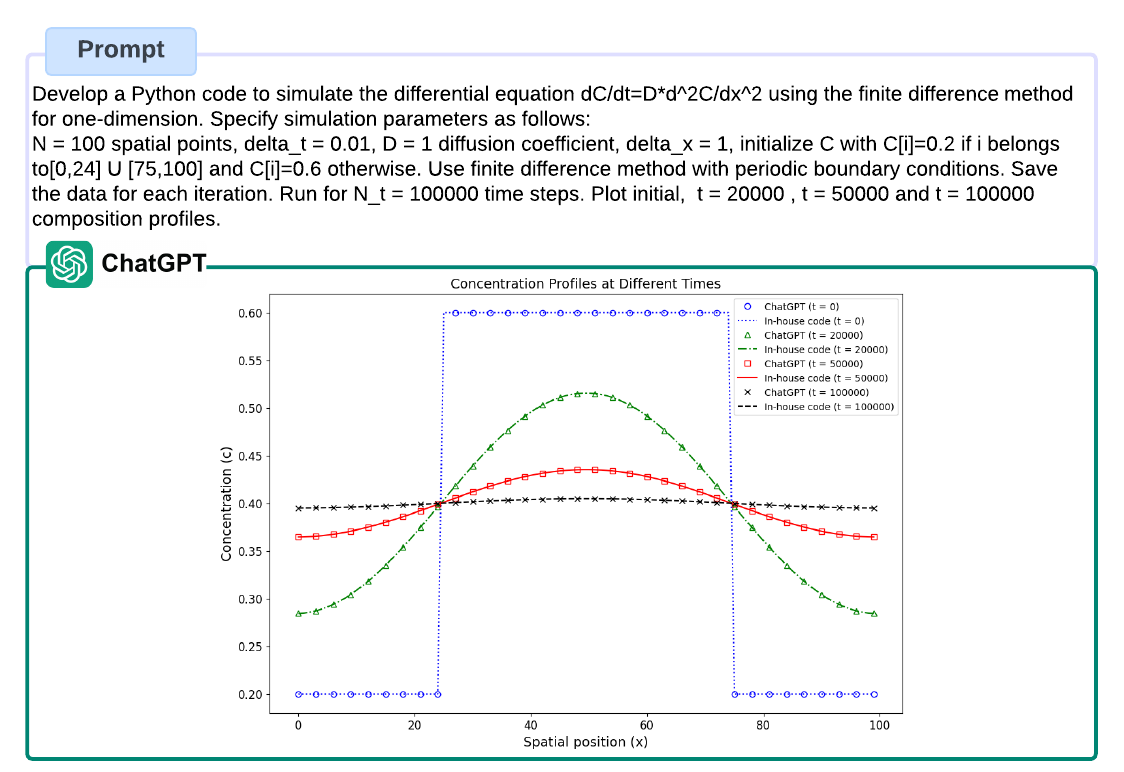}
\caption{ (Color online) As a first case study (Sec.~\ref{sec:task1}), we depict the ChatGPT response generated after entering the given prompt in a zero-shot manner directly in its AI chatbot. The time evolution of the composition field generated by ChatGPT is compared with excellent accuracy with the in-house implementation of the identical problem. Over time, the initial solute concentration profile (step function) approaches the equilibrium value ($c = 0.4$) due to diffusion.}\label{fig:ficks}
\end{figure}

Diffusive phase transformations and associated microstructure evolution involve long-range diffusion. Therefore, we begin the first example with a very simple yet well-known PDE, the transient solute diffusion equation given by Fick's second law (mass conservation). In this particular example, we focus on the accuracy and efficiency of ChatGPT implementation. We consider the following form of Fick's second law of diffusion~\cite{callister,porter_book} in 1D,
\begin{equation}\label{eq:diffusion}
\frac{\partial c(x, t)}{\partial t} = D \frac{\partial^2 c(x, t)}{\partial x^2},
\end{equation}
where $c(x, t)$ is the concentration profile at a position $x$ and time $t$, and $D$ is the diffusivity of solute. We solve on a domain of $x \in [0,\ 99]$, assuming the initial concentration as a step function between 0.6 ($24<x<75$) and 0.2 (otherwise) and periodic boundary condition. We use $\Delta x = 1.0$, $\Delta t = 0.01$, $D =1$, and $N_t = 100000$ time steps. We use an iteration scheme using the explicit Euler method in time and a second-order central finite difference scheme in space. To solve Eq.~\eqref{eq:diffusion}, we input the PDE and the simulation setup and parameters in the ChatGPT dialogue box using a single prompt, as shown in Fig.~\ref{fig:ficks}. Then ChatGPT performs the necessary calculations and analysis, that is, describes its understanding of the algorithmic program steps, writes the code in Python, executes the code, and finally generates the output plot of $c(x, t)$ (using Matplotlib). We show the given prompt and the ChatGPT output in Fig.~\ref{fig:ficks}. When compared with our in-house implementation of the identical problem, ChatGPT produces accurate execution results. Note the use of question prompt in a zero-shot manner and that too formatted in natural language, demonstrating efficiency in the ChatGPT approach.
 
\subsubsection{General observations}
We note that ChatGPT works perfectly when the explicit form of the PDE is given, as expected, and also when queried as Fick's second law. Besides Python, it can support programming in many languages, including Matlab/Octave, C/C++, Java, and R; however, as per ChatGPT, ``it can't directly compile or execute these codes or visualize the plots in its environment.'' Importantly, it can autonomously learn from an incorrect code that it writes and then be able to ``correct the code and run the simulation again'' by itself. It should be noted that for repeated query of the same prompt may generate different code structures. The prompt does not need to be case-sensitive. Importantly, ChatGPT is \textit{context-aware}, meaning, even if a variable is defined differently at various places in the prompt, it understands the context of the variable and considers them as the same. It extensively uses NumPy, SciPy, Matplotlib, scikit-learn, and other Python libraries and wrappers as required for coding and data analysis; and it can save the output data and images in arbitrary formats. Although not shown here, ChatGPT is able to solve Eq.~\eqref{eq:diffusion} correctly using the Fourier spectral method~\cite{fftw} and implicit time-stepping schemes. Also, it is somewhat capable of modifying the code that can be run on external parallel CPU and GPU clusters using OpenMP~\cite{Openmp}, MPI~\cite{Mpich}, and CUDA~\cite{Cuda}. As we will show later, ChatGPT can efficiently perform 2D simulations on reasonably-sized domains; however, as per ChatGPT, ``3D simulations might be constrained to domains with dimensions in the order of hundreds to a few thousand grid points along each axis depending on the complexity of the problem.'' Importantly, ChatGPT allows the user to copy the code to perform large-scale simulations on external systems. It is able to give reasonable material-specific data after a literature review (e.g., $D \approx 7 \times 10^{-9}$) when we ask: ``what is the diffusion coefficient for 3 wt\% Cu in Al at 877 K?'', noting that repeating the same query generated a more accurate answer ($D \approx 2 \times 10^{-9}$)~\cite{farzadi2008}. However, it is unable to provide this data for many other material systems but recommends correct online resources where to look for to obtain these parameter values. Also, when asked for more technical details like, ``is the above analysis valid when $D$ is a function of $c$?'', it correctly responded as ``no'' and provided the conceptual modifications required in the simulation code. These are the general observations we had while interacting with ChatGPT; hence, for the sake of brevity, we do not repeat the above queries for other case studies henceforth. However, depending on the complexity of the PDE requested to solve by the user, the above specific observations may vary from task to task. 

\subsection{Task 2: Cahn-Hilliard equation}\label{sec:task2}
The second case study corresponds to material problems involving long-range diffusion, as seen in microstructure evolution \textit{via} spinodal decomposition or phase separation in general~\cite{ghosh2017pccp,ghosh2020chemical,ghosh2024_fractal}. We use the Cahn-Hilliard PDE~\cite{Cahn1958} (hyper-diffusion equation with fourth-order derivatives of $c$) to simulate spinodal decomposition in a binary alloy of two-phase systems,
\begin{equation}\label{eq:ch}
\frac{\partial c (r, t)}{\partial t} =  \nabla \cdot M \nabla \left( \frac{\partial f(c)}{\partial c} - \kappa_c \nabla^2 c\right),
\end{equation}
where $r$ is the position, $M$ is the mobility, $f$ is the chemical free energy of the system, and the second term is the interfacial free energy of the diffuse interface, in which $\kappa_c$ is the gradient energy coefficient. Equation~\eqref{eq:ch} is based on the total free energy of the given material given by,
\begin{equation}\label{eq:functional1}
    \mathcal{F} (c, \nabla c) = \int_v \left[ f(c) + \frac{\kappa_c}{2} |\nabla c|^2 \right] \, \, dv,
\end{equation}
which is the starting point for most phase-field models~\cite{ghosh2024_fractal,ghosh2022tusas}, but there can be additional terms due to the coupling of the additional physics, as we will explore this in subsequent case studies. We consider $f(c) = A_cc^2(1-c)^2$, a double-well potential whose wells define the phases. We solve Eq.~\eqref{eq:ch} on a $128 \times 128$ domain with $\Delta x = \Delta y = 1.0$, $\Delta t = 0.1$, $\kappa_c = 2$, $M = 0.1$, $A_c = 1$, $N_t = 40000$, and periodic boundary conditions in all directions. We assume a random noise of variance $\pm$ 0.01 that is added to the initial matrix composition ($c_0 = 0.5$). We implement with the explicit Euler finite-difference algorithm. 

\begin{figure}[htbp]
\includegraphics[width=\textwidth]{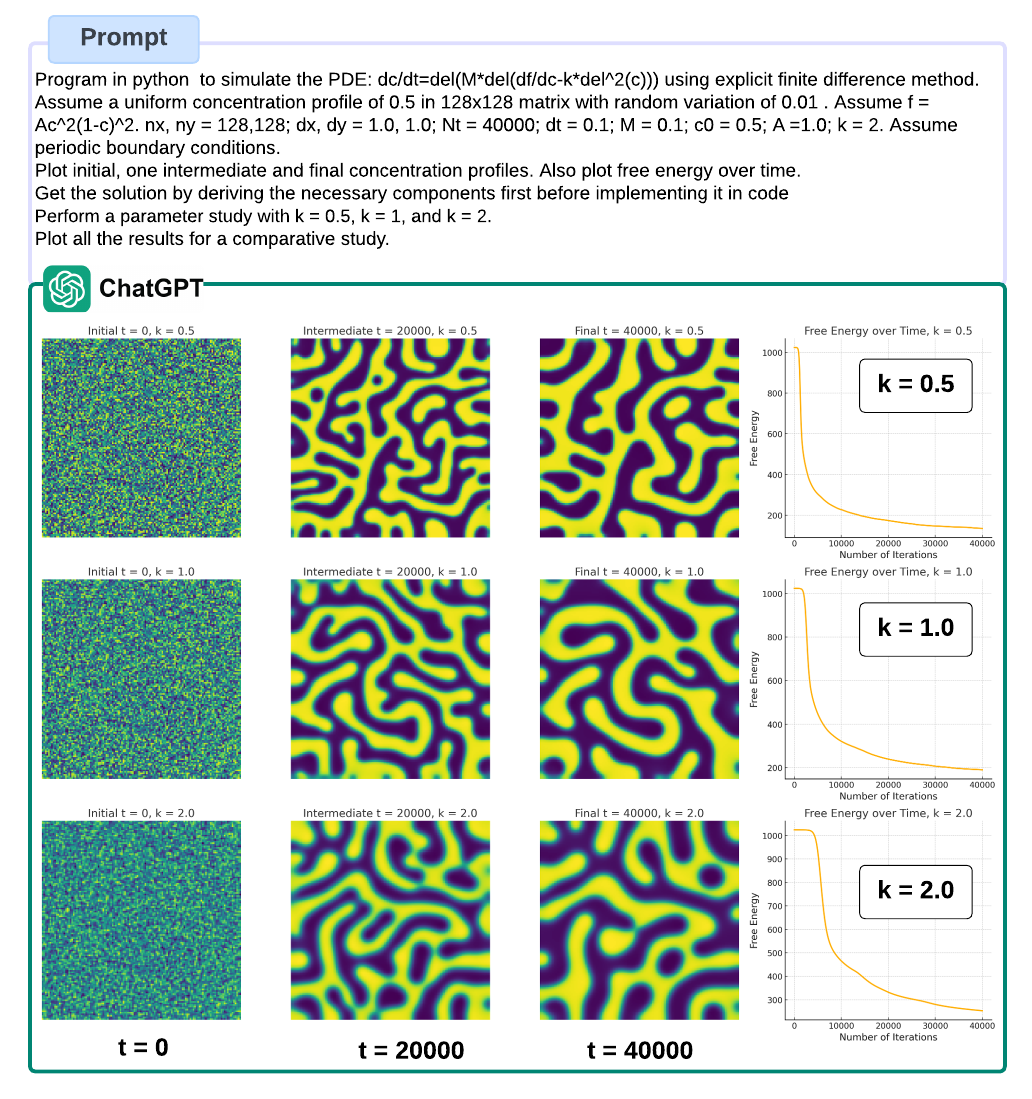}
\caption{(Color online) As a second case study (Sec.~\ref{sec:task2}), we depict the ChatGPT response generated after entering the given prompt in a zero-shot manner directly in its AI chatbot. The time evolution of the composition field in the simulation box leads to phase separation of the binary alloy into $A$-rich (red) and $B$-rich (blue) phases. The coarsening of the microstructure phases over time ($t$) is evident. As expected, the total free energy of the system decreases with time (subplots in the right panel). We ask ChatGPT to perform a parameter study with varying values of the gradient energy coefficient $k$. The output in terms of microstructure and free energy evolution over time is shown in the top row for $k$ = 0.5, middle row for $k$ = 1.0, and bottom row for $k$ = 2.0. The interface becomes more diffuse with increasing value of $k$, promoting faster coarsening of phases to reduce the interfacial (and hence total) free energy of the phase-separating system.}\label{fig:ch}
\end{figure}

The corresponding prompt and the ChatGPT output are shown in Fig.~\ref{fig:ch}. The general observations are as follows. With time, the initial binary matrix phase separates into $A$-rich and $B$-rich phases, followed by the coarsening of the phases (i.e., Ostwald ripening)~\cite{steinbach2013phase,voorhees1985}. The phase separation and coarsening processes are direct consequences of the reduction in total free energy of the system (right subpanel in Fig.~\ref{fig:ch}) owing to a significant reduction in the amount of interfaces in the microstructure~\cite{lee2013comparison}. Although not shown here, we confirm the accuracy of the output against our in-house code simulation of the identical problem. 

Next, we test the capability of ChatGPT to perform parameter studies. For reference, the parameter $k$ controls the interfacial energy between different phases in the system, hence critical for phase separation dynamics. We ask ChatGPT to run  Cahn-Hilliard simulation (Eq.~\eqref{eq:ch}) for three different $k$ values, as demonstrated in the prompt given in Fig.~\ref{fig:ch}. For each value of $k$, ChatGPT calculates the microstructure evolution over time and the corresponding reduction in the free energy of the system. ChatGPT clearly captures the effect of increasing value of $k$: compared to systems with small $k$, large $k$ values lead to faster diffusion and more diffuse interfaces, which promote faster coarsening of phases to reduce the total free energy of the system more rapidly. Since such a general parameter study can be performed in a straightforward manner by providing different values of the parameter of choice in the prompt, henceforth, we do not repeat such a study for simplicity. 

Due to severe time step restrictions (due to the $\nabla^4 c$ term in Eq.~\eqref{eq:ch}), ChatGPT is unable to work on complex domain geometries, including 3D and non-rectangular 2D, and even on square geometry for large-scale simulations. When we ask it to simulate for more time steps, it responded, ``it took too long and exceeded the time limit for processing.'' Also, when asked for materials-specific values for $\kappa_c$ and $M$, it provided general values, $\kappa_c$ $\approx$ $10^{-11}$ to $10^{-9}$ J/m and $M$ $\approx$ $10^{-9}$ to $10^{-12}$ m$^2$/s, but not for any specific alloy, likely due to lack of domain-specific knowledge in the training data set. We note that repeated queries of the same prompt may be needed to produce the desired results and plots. Further, when we ask ChatGPT about a causal reasoning question like ``what is the relationship between $k$ and equilibrium'', it correctly responds: ``a larger $k$ value increases the energy penalty for sharp concentration gradients, which drives the system to smooth out these gradients more rapidly and thus leads to a faster approach to equilibrium.'' Finally, the ChatGPT code can be run on the client-side system to simulate large-scale, long-time microstructure evolution process with complex domain geometries.

\subsection{Task 3: Coupled Cahn-Hilliard and Ginzburg-Landau or Allen-Cahn equations}\label{sec:task3}
Next, we couple generalized PDEs of Ginzburg-Landau or Allen-Cahn~\cite{Allen1979} and Cahn-Hilliard~\cite{Cahn1958} type to study the coupled physics of conserved (concentration: $c$) and non-conserved (order parameter: $\phi$) parameters during microstructure evolution. The general approach could be used to model curvature-driven problems (e.g., grain growth and coarsening) or order-disorder transformations~\cite{provatasbook}. The corresponding PDE for $c$ is given by Eq.~\eqref{eq:ch}; and for $\phi$ we consider,
\begin{equation}\label{eq:ca}
\frac{\partial \phi (r, t)}{\partial t} =  - L \left( \frac{\partial f}{\partial \phi} - \kappa_\phi \nabla^2 \phi\right),
\end{equation}
where $L$ is a mobility parameter, $f$ is a double-well free energy potential, and $\kappa_\phi$ is the gradient energy coefficient. For simplicity, we use $f (c, \phi) = A_c c^2(1-c)^2 + A_\phi \phi^2(1-\phi)^2$ that couples $\phi$ and $c$. In terms of solution complexity, the coupled system contains a $\nabla^2\phi$ term in addition to the usual $\nabla^2c$ and $\nabla^4c$ terms in Eq.~\eqref{eq:ch}. In this case, the free energy of the material is given by,
\begin{equation}\label{eq:functional2}
    \mathcal{F} (c, \phi, \nabla c, \nabla \phi) = \int_v f(c, \phi) + \frac{\kappa_c}{2} |\nabla c|^2 + \frac{\kappa_\phi}{2} |\nabla \phi|^2 \, \, dv,
\end{equation}
which makes the starting point for many advanced phase-field models~\cite{ghosh20183d,ghosh2017_eutectic}.

We solve coupled PDEs, Eq.~\eqref{eq:ch} and~\eqref{eq:ca}, on a $128 \times 128$ domain with $\Delta x = \Delta y = 1.0$, $\Delta t = 0.01$, $\kappa_c = \kappa_\phi = 2$, $M = 0.1$, $L = 0.1$, $A_c = 1$, $A_\phi = 2$, $N_t = 20000$, and periodic boundary conditions in all directions. We assume an initial circular grain (radius of 20 grids, $\phi = 1$ and $c = 1$) embedded into a large second grain ($\phi = 0$ and $c = 0$). We implement the system with an explicit Euler finite-difference algorithm. Since our objective is to test the capability of ChatGPT in solving coupled PDEs with additional physics, we keep the above model parameters non-dimensional and numerically affordable. 

\begin{figure}[ht]
\includegraphics[width=\textwidth]{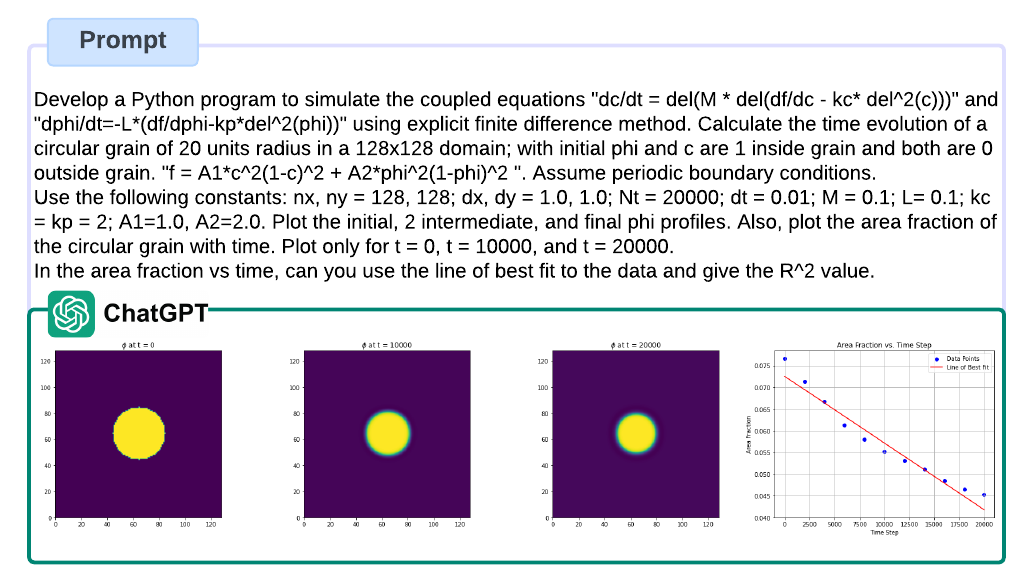}
\caption{(Color online) As a third case study (Sec.~\ref{sec:task3}), we depict the ChatGPT response generated after entering the given prompt in a zero-shot manner directly in its AI chatbot. For reference, we only show the time evolution of the $\phi$ field. The circular grain shrinks over time due to the curvature-induced diffusion effects. The area fraction of this grain decreases linearly with time, with the $R^2$ of a line of best fit $\approx 0.957$.}\label{fig:cha}
\end{figure}

The corresponding prompt and the ChatGPT response are shown in Fig.~\ref{fig:cha}. The initial interface becomes diffuse with time, and the circular grain continues to shrink over the time of the simulation, as expected. The ChatGPT is able to characterize such shrinkage behavior of the grain following the change in its radius with increasing time, as evident from Fig.~\ref{fig:cha} (bottom right panel). We observe the identical grain behavior using our in-house code. Note that due to the increased complexity of the coupled system, it takes a longer time for each iteration when compared to that with Eq.~\eqref{eq:ch}, thus reducing the timescale limit that ChatGPT can process (i.e., 40000 in Task 3 vs. 20000 in Task 4). To further speed up the simulations, for example, GPU-parallel resources are required. Unfortunately, as per ChatGPT ``the environment here does not support direct execution of GPU-accelerated code, as it requires access to physical GPU hardware and specific software libraries that are not available in this setup.'' Also, when asked to ``fit the area fraction vs. time curve using a line of best fit and provide the $R^2$ value'', it correctly estimated the linear relationship~\cite{hillert1965theory} between the time and area fraction with $R^2 \approx 0.957$, demonstrating basic data analysis ability of ChatGPT ``on-the-fly.'' Furthermore, as a query for domain knowledge, ``why is the area fraction decreasing with time?'', it correctly recognized the critical role of curvature-driven interface migration and the Gibbs-Thomson effect~\cite{porter_book}. We note that for more complex model formulations and parameters, such as polycrystalline microstructure evolution due to grain growth~\cite{fan1997computer}, ChatGPT should be able to generate the code, which can then be executed with external software.

We must admit that without specific numerical instructions, ChatGPT and LLMs in general produce different code structures with different numerical solution schemes (in terms of different Python functions, optimizations, array loop structures, boundary condition implementations, etc.). Hence, the stability and accuracy of the resulting code will differ for each output generation. Consequently, the quantitative details of the output will vary, although the simulation code may be correct, as we note in this particular case. While simulating the same problem using our in-house code, we observe a $R^2 \approx 0.923$ compared to ChatGPT output with $R^2 \approx 0.957$. Such differences in the quantitative measure of the simulation results vary depending on the specific implementation schemes ChatGPT uses to produce the output, noting that the solution schemes may differ upon repeated user requests even for solving the same problem.

\subsection{Task 4: Solidification and dendritic growth}\label{sec:task4}
In materials solidification, the most common microstructures are made of dendrite crystals~\cite{kurz2019progress,kurz2021progress}. Compared to previous case studies, this case study involves additional physics, including anisotropic diffusion, interfacial energy, and solidification, among others. We use the famous Kobayashi~\cite{kobayashi1993} phase-field formulation (based on Ginzburg-Landau or Allen-Cahn Eq.~\eqref{eq:ca}) to simulate dendritic growth in a pure material, noting that this model makes the basis for more complex dendrite models derived later~\cite{Echebarria2004,Trevor2017}. We solve two coupled PDEs describing the time evolution of phase-field $\phi$ and temperature $T$ as,
\begin{align}\label{eq:phit}
 \tau \frac{\partial \phi}{\partial t}   = \frac{\partial}{\partial y} \left(\epsilon \frac{\partial \epsilon}{\partial \theta} \frac{\partial \phi}{\partial x} \right)  - \frac{\partial}{\partial x} \left( \epsilon \frac{\partial \epsilon}{\partial \theta} \frac{\partial \phi}{\partial y)}\right)  + \nabla \cdot\left(\epsilon^2 \nabla \phi \right) + \phi (1 - \phi) (\phi - \frac{1}{2} + \frac{\alpha}{\pi} \arctan \left[ \gamma (T_m - T) \right]),
\end{align}
and
\begin{equation}\label{eq:tempt}
\frac{\partial T}{\partial t} = \nabla^2 T + K \frac{\partial \phi}{\partial t}.
\end{equation}
Here, $\tau$ is the relaxation time; $\epsilon(\theta) = \bar{\epsilon}\left[1 + \delta \cos(m(\theta-\theta_0))\right]$ is a $m$-fold crystal anisotropy function with $\theta = \arctan \left( \frac{\partial \phi / \partial y}{\partial \phi /\partial x} \right)$ the interface orientation angle and $\delta$ the anisotropy strength; $\alpha$ and $\gamma$ are positive constants; $T_m$ is the melting temperature; and $K$ is the latent heat. Although not shown here for brevity, a Ginzburg-Landau type system free energy (similar to Eq.~\eqref{eq:functional2}) is assumed to obtain Eqs.~\eqref{eq:phit} and~\eqref{eq:tempt}.

We solve coupled PDEs, Eqs.~\eqref{eq:phit} and ~\eqref{eq:tempt}, assuming a solid ($\phi = 1$) circular seed of radius of 5 grid units at the center of a 2D $200 \times 200$ domain filled with liquid ($\phi = 0$). We use $\Delta x = \Delta y = 0.03$, $T = 0$, $\Delta t = 0.0001$, $\tau = 0.0003$, $\bar{\epsilon} = 0.01$, $\delta = 0.02$, $m = 6$, $\theta_0 = 0.05$, $\alpha = 0.9$, $T_m = 1.0$, $\gamma = 10.0$, $K = 2.0$, $N_t = 2500$ time steps, and periodic boundary conditions for both $\phi$ and $T$ in all directions. We use the explicit time marching scheme with a central finite difference method to discretize the PDEs. 

\begin{figure}[htbp]
\includegraphics[width=\textwidth]{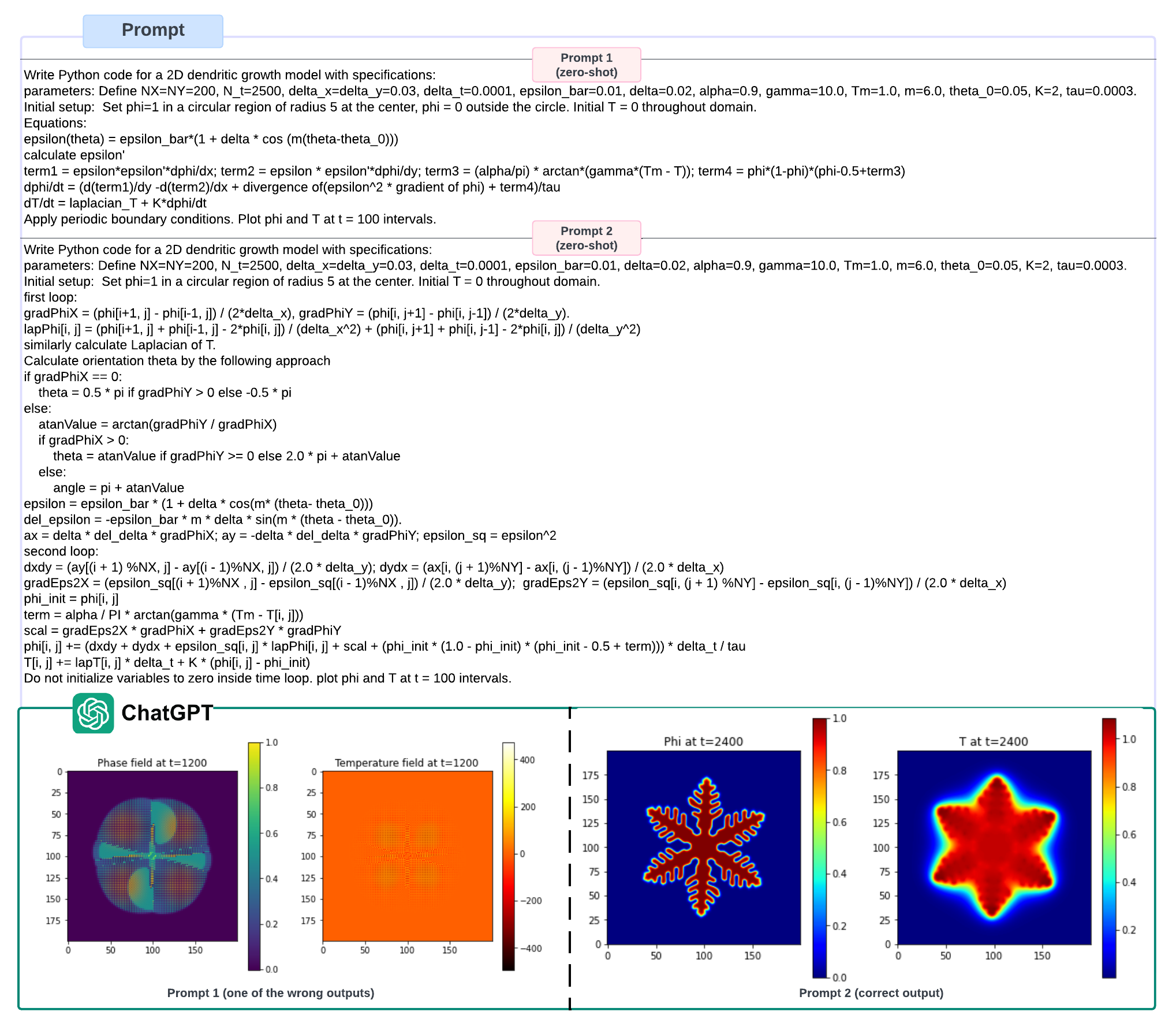}
\caption{(Color online) As a fourth case study (Sec.~\ref{sec:task4}), we depict the ChatGPT-4 response after entering the prompts in a zero-shot manner in the AI chatbot. Prompt 1 produces the wrong code and output. Since ChatGPT produces different codes for repeated user calls with the same prompt, this is just one of the wrong outputs that we present for reference. Prompt 2 produces the correct code and output. Using Prompt 2, we show the dendrite growth at time $t = 2400$ by plotting the $\phi$ and $T$ fields. The six primary arms and the well-developed secondary branches can be clearly seen. These results perfectly match with the in-house code execution results for the identical problem.}\label{fig:dendrite}
\end{figure}

The corresponding prompt and the ChatGPT response are shown in Fig.~\ref{fig:dendrite}. After providing the equation and parameters in a prompt formatted in natural language to ChatGPT (Prompt 1 in Fig.~\ref{fig:dendrite}), it generates an executable yet logically incorrect code and thus fails to produce the correct output (bottom left subpanel in Fig.~\ref{fig:dendrite}). In particular, it could not handle the derivatives involving anisotropy (first two terms in Eq.~\eqref{eq:phit}) and thus could not resolve the dependency between these equations \textit{via} $\partial \phi/\partial t$ in Eq.~\eqref{eq:tempt}. Therefore, we had to provide more detailed technical instructions (inspired by our in-house code of the same problem) on how to solve the coupled PDEs. We update the prompt with specific \textit{rules} on (a) how to calculate the derivatives and Laplacian using a five-point stencil; (b) how to work on the explicit forms of the derivatives, particularly, involving with the complex anisotropy terms in Eq.~\eqref{eq:phit}; (c) how to code the spatial derivatives of anisotropy with respect to $\theta$, and (d) how to explicitly code the iteration loops to correctly calculate the time derivatives. With all the extensive feedback provided \textit{via} a series of prompt iterations, so-called ``Chain-of-Thought'' (CoT) prompting, ChatGPT was able to learn from these interactions and correct errors by itself, eventually generating the correct code (see supplementary material). We present these series of prompt iterations in a zero-shot manner in Fig.~\ref{fig:dendrite} (Prompt 2). Due to the complex nature of the PDEs and severe time restriction ($\Delta t = 10^{-4}$), processing of the code takes a long time; hence, ChatGPT was unable to execute the code in its environment. Thus, we export this code to an external system and execute it locally to obtain the solution output. The output $\phi$ and $T$ profiles are shown in Fig.~\ref{fig:dendrite}, where the initial circular solid seed (not shown) grows with time to produce dendritic structures with six primary arms with well-defined side branches. 

It is evident that several expert-driven iterations and feedback need to be given to solve multi-physics coupled PDEs in ChatGPT. Also, how much information needs to be input to generate accurate code/output that remains a challenge. We note that since LLMs are essentially a text-completer~\cite{geron2022}, the user should input nearly all the intricate code tasking details, as if like ``forcing'' the model to generate the desired output, as evident in this particular case. Nevertheless, this approach will almost take away the convenience of the ChatGPT approach, as seen in previous case studies (Sec.~\ref{sec:task1}-\ref{sec:task3}). Thus, it will be more prudent to write the coding task from scratch or build it on top of the code base provided by ChatGPT for a better investment of time and skills. 

\section{Summary and Outlook} \label{sec:summary}
In this work, we test the capability of ChatGPT to solve several well-known PDEs central to materials science. The solution of these PDEs, under appropriate simulation setup and parameters, provides the spatial and temporal dependence of the microstructure variables in metallic alloys. To this purpose, we use four benchmark case studies for numerical implementations of phase-field model PDEs with increasing order of complexity in terms of additional physics. In this context, our objective is to replace programming and compiling with prompt writing how far we can explore ChatGPT for generating the correct solutions of the input PDEs, noting that there is a significant potential for coding skills to transition fully to natural language in the future. From this work, we highlight the following observations and remarks:

\begin{itemize}

\item ChatGPT can be effectively leveraged to address a range of tasks aiding materials research, such as writing and explaining scientific computer codes, running Python codes, generating visualization files, performing data analysis, and even writing scripts for running simulations on mainstream software such as MATLAB~\cite{Matlab}. As per our experience (Table~\ref{table:summary}), other LLMs are not so robust yet when it comes to scientific computing and data analysis. Thus, many of the latest conversational AI chatbots, such as Copilot~\cite{copilot}, are powered by GPT-4 Turbo~\cite{gpt4_turbo}, the latest iteration in the GPT models (as of April 1, 2024).

\item In particular, ChatGPT is seemingly professional at generating multi-physics codes, with its particular strength in scientific computing in Python. However, we do not recommend executing the code on ChatGPT due to its processing constraints on simulation length and time scales (see Sec.~\ref{sec:task4}). With increasing problem complexity, we note that repetition of the same query is often needed to produce the desired output. The dependence on reliable internet speed aggravates the situation even further. Fortunately, ChatGPT allows the user to copy the code to use in external software systems. 

\item In case of complex PDEs with interacting many physical parameters (Sec.~\ref{sec:task4}), ChatGPT is likely to introduce inaccuracies in the code and thus fails to produce the full solution process, indicating the computing constraints and its low capability to complex PDE solution. Thus, additional, specific details must be provided based on the complexity of the coding task. However, the risk remains for more hidden errors and output inconsistencies, which one can often experience when working with ChatGPT. Thus, expert-driven fact-checking and verification processes become essential. Although ChatGPT is capable of providing a code base as a good starting point to tackle such complex tasks, a meticulous amount of coding tasks still remains at hand. 

\item We find that the performance of ChatGPT significantly decreases as the complexity of PDEs increases. This is due to the lack of domain knowledge that aggravates the gap between its understanding of the problem and the real situation. With the increasing complexity of PDEs, we note that it is becoming more of a trial-and-error approach to solve PDEs and less of a scientific approach. Thus, ChatGPT is not yet efficient at handling domain-specific specialized research involving complex PDEs for which an in-depth understanding of the given problem is required, as shown in Sec.~\ref{sec:task4}. However, with more technical advancements of these models in the next iterations of development (e.g., GPT-5), they should be able to handle increasingly complex PDEs. 

\item It is clear from our work (Sec.~\ref{sec:task1}-\ref{sec:task3}) that ChatGPT shows acceptable performance in exploring general scientific tasks, that is, to solve PDEs up to a certain complexity. In this regard, the ability of ChatGPT can be further improved to some extent \textit{via} chain prompting by providing iteration and ``intelligent'' feedback to ChatGPT. These aspects will be indeed helpful for classroom teaching and training of new materials researchers where mostly low-to-medium level complex queries are addressed.

\item ChatGPT can fail spectacularly in generating code that does not work. For the most difficult problem to solve in this study (i.e., Task 4 for dendrite crystal growth), ChatGPT fails to produce the correct code. We carefully evaluate the errors in the generated codes after making repeated calls with the same prompt in ChatGPT. However, the errors are not consistent as they differ in different codes. We note that typical errors range from implementation errors in the initial and boundary conditions, logical errors in computing derivatives and implementing array loops, broken code due to the timeout limit in ChatGPT, and sometimes ignored instructions in the input/output. Since showing all the wrong outputs is impractical, we show a few of these in the supplementary material for demonstration. In Sec.~\ref{sec:task4}, we discuss the error mitigation strategies to produce the desired output in the context of the dendrite growth problem. However, solving dendrite growth in ChatGPT is not trivial because of the large-sized prompt, and there is a limit to how many characters within a chat ChatGPT will ``remember'' so it will frequently make mistakes due to losing the context of early prompts or its early answers and sometimes generate a broken code due to the timeout limit. Therefore, we attempt to use a single prompt that will minimize inference time and reduce the computational cost, improving the effectiveness of the ChatGPT approach in error mitigation. Note that evaluating errors and improving the performance of ChatGPT is a broad topic, which is not the focus of this work, but it makes an interesting future study. Remarkably, the newest and latest GPT model, ChatGPT-4o~\cite{gpt4o}, leads to better results for coding tasks to simulate the dendritic growth problem even using a single prompt (prompt 1 in Fig.~\ref{fig:dendrite}, see supplementary for details). The ChatGPT-4~\cite{gpt4} could not produce the correct code and output using the same prompt (see left subpanel in Fig.~\ref{fig:dendrite}).

\item We show that LLMs in general are useful in coding tasks. Moreover, even if the generated code produces the correct output, the accuracy of the results may not be useful for the targeted applications. For instance, the simulation accuracy required for materials education purposes is different from that required in critical applications such as designing materials for healthcare and aerospace, where accuracy and reliability are paramount. Perhaps more troubling is that the LLM algorithms are ``black boxes''; thus, it is difficult to know why they are generating the code that they do, and there is no measure of uncertainty associated with those outputs. Thus, uncertainty quantification (UQ) may be essential in the future to enhance the reliability and trustworthiness of LLMs, fostering greater acceptance and utilization in practical applications.

\item Finally, every reward comes with a cost. ChatGPT will undoubtedly reduce the need for certain skills; most importantly, scientific coding can be accomplished more efficiently with reduced time investment. Thus, it remains to be seen how the use of LLMs challenges traditional ways of education and training of researchers. Most importantly, beginners must remain cautious with answers from ChatGPT and in relying heavily on it, as it may hinder the development of their research skills. Also, the challenge remains to be seen in terms of the extent to which the research process should be outsourced to LLMs and what academic skills remain essential to researchers. On the other hand, these AI tools will eventually open up new frontiers in various domains and introduce new skills, such as prompt engineering and generative AI in general, which may significantly boost research productivity. However, with increasingly complex problems, more expert guidance is required for LLMs to perform the task accurately. Thus, researchers should gain sufficient expertise in their domain knowledge before they can leverage ChatGPT to expedite their workflow. 
\end{itemize} 

There is no denying that LLMs have the potential to impoverish our own writing and thinking skills since often we have to do only a little to obtain a sophisticated text or code~\cite{schulze2024empirical}. While the debate will continue regarding its implication on scientific practice and responsible use of LLMs for research and the security risk it poses, the benefit-to-risk ratio remains very high~\cite{van2023chatgpt} and is improving with each passing day. For example, ChatGPT-4 now cites the sources for its answers, addressing some of the plagiarism issues inherent to LLM-based applications (inherited from training data)~\cite{tweet}. With the rapid increase in model ``parameters'' and ``tokens'' (Table~\ref{table:llm}) and collaborative synergy with human expertise, LLMs offer a promising outlook for accelerating materials research and exploration that seem out of reach today. Consequently, the enhanced proficiency of LLMs in materials modeling at different scales could enable them to evolve into AI-powered ICME agents in the future.
\section*{Acknowledgments}
S. Ghosh acknowledges the support from FIG (SRIC office, IIT Roorkee) and SERB (Government of India).
\section*{Supplementary Material}
We provide the Python codes generated by ChatGPT as a response to the given prompts for four tasks, as depicted in Figs.~\ref{fig:ficks}-\ref{fig:dendrite}. These files are named as \texttt{task1.py}, \texttt{task2.py}, \texttt{task3.py}, and \texttt{task4.py}. We share the ChatGPT chat history to solve the diffusion problem (Task 1) in \texttt{Task1-chatgpt.pdf}. We share the ChatGPT workspace for dendritic growth problem (Task 4) with incorrect output in \texttt{Task4-wrong.pdf}. The output from ChatGPT-4o for Task 4 is given in \texttt{Task4-chatgpt4o-output.pdf}. 

\section*{Conflict of Interest Statement}
The authors have no conflicts to disclose.

\section*{Data Availability Statement}
The data that supports the findings of the study are available from the corresponding author upon reasonable request.

\begin{thebibliography}{}
\expandafter\ifx\csname url\endcsname\relax
  \def\url#1{\texttt{#1}}\fi
\expandafter\ifx\csname urlprefix\endcsname\relax\def\urlprefix{URL }\fi
\expandafter\ifx\csname href\endcsname\relax
  \def\href#1#2{#2} \def\path#1{#1}\fi

\end{thebibliography}


\begin{thebibliography}{10}
\expandafter\ifx\csname url\endcsname\relax
  \def\url#1{\texttt{#1}}\fi
\expandafter\ifx\csname urlprefix\endcsname\relax\def\urlprefix{URL }\fi
\expandafter\ifx\csname href\endcsname\relax
  \def\href#1#2{#2} \def\path#1{#1}\fi

\bibitem{van2023chatgpt}
E.~A. Van~Dis, J.~Bollen, W.~Zuidema, R.~Van~Rooij, C.~L. Bockting, {ChatGPT}:
  five priorities for research, Nature 614~(7947) (2023) 224--226.
\newblock \href {http://dx.doi.org/https://doi.org/10.1038/d41586-023-00288-7}
  {\path{doi:https://doi.org/10.1038/d41586-023-00288-7}}.

\bibitem{schulze2024empirical}
L.~Schulze~Balhorn, J.~M. Weber, S.~Buijsman, J.~R. Hildebrandt, M.~Ziefle,
  A.~M. Schweidtmann, {Empirical assessment of {ChatGPT}'s answering
  capabilities in natural science and engineering}, Scientific Reports 14~(1)
  (2024) 4998.
\newblock \href {http://dx.doi.org/https://doi.org/10.1038/s41598-024-54936-7}
  {\path{doi:https://doi.org/10.1038/s41598-024-54936-7}}.

\bibitem{geron2022}
A.~G{\'e}ron, {Hands-on machine learning with Scikit-Learn, Keras, and
  TensorFlow}, O'Reilly Media, Inc., Newton, MA, 2022.

\bibitem{vaswani2017attention}
A.~Vaswani, N.~Shazeer, N.~Parmar, J.~Uszkoreit, L.~Jones, A.~N. Gomez,
  L.~Kaiser, I.~Polosukhin, Attention is all you need, in: Advances in neural
  information processing systems, 2017, pp. 5998--6008.

\bibitem{commoncrawl}
{Common Crawl}, \url{https://commoncrawl.org/} (2024).

\bibitem{wikipedia}
{Wikipedia}, \url{https://www.wikipedia.org/} (2023).

\bibitem{alto2023modern}
V.~Alto, Modern Generative AI with {ChatGPT} and OpenAI Models: Leverage the
  capabilities of OpenAI's LLM for productivity and innovation with GPT3 and
  GPT4, Packt Publishing, Birmingham, UK, 2023.

\bibitem{lei2024materials}
G.~Lei, R.~Docherty, S.~J. Cooper, Materials science in the era of large
  language models: a perspective, arXiv preprint arXiv:2403.06949\href
  {http://dx.doi.org/https://doi.org/10.48550/arXiv.2403.06949}
  {\path{doi:https://doi.org/10.48550/arXiv.2403.06949}}.

\bibitem{liu2023generative}
Y.~Liu, Z.~Yang, Z.~Yu, Z.~Liu, D.~Liu, H.~Lin, M.~Li, S.~Ma, M.~Avdeev,
  S.~Shi, Generative artificial intelligence and its applications in materials
  science: Current situation and future perspectives, Journal of Materiomics
  9~(4) (2023) 798--816.
\newblock \href {http://dx.doi.org/https://doi.org/10.1016/j.jmat.2023.05.001}
  {\path{doi:https://doi.org/10.1016/j.jmat.2023.05.001}}.

\bibitem{llmlingua}
{{LLMLingua-2}}, \url{https://llmlingua.com/} (2023).

\bibitem{gpt3}
{GPT-3}, \url{https://openai.com/blog/gpt-3-apps/} (2021).

\bibitem{gpt4}
{GPT-4}, \url{https://openai.com/gpt-4/} (2023).

\bibitem{grok}
{Grok}, \url{https://grok.x.ai/} (2023).

\bibitem{gemini}
{Gemini}, \url{https://gemini.google.com/app/} (2024).

\bibitem{llama}
{Llama 2}, \url{https://llama.meta.com/} (2023).

\bibitem{copilot}
{Copilot}, \url{https://copilot.microsoft.com/} (2023).

\bibitem{cluade}
{Cluade-3}, \url{https://claude.ai/} (2024).

\bibitem{mistral}
{Mistral AI}, \url{https://mistral.ai/} (2023).

\bibitem{frieder2024mathematical}
S.~Frieder, L.~Pinchetti, R.-R. Griffiths, T.~Salvatori, T.~Lukasiewicz,
  P.~Petersen, J.~Berner, Mathematical capabilities of {ChatGPT}, Advances in
  Neural Information Processing Systems 36.

\bibitem{koceska2023can}
N.~Koceska, S.~Koceski, L.~K. Lazarova, M.~Miteva, B.~Zlatanovska, Can chatgpt
  be used for solving ordinary differential equations, Balkan Journal of
  Applied Mathematics and Informatics 6~(2) (2023) 103--114.

\bibitem{orlando2023assessing}
G.~Orlando, Assessing chatgpt for coding finite element methods, Journal of
  Machine Learning for Modeling and Computing 4~(2).

\bibitem{dave2023chatgpt}
T.~Dave, S.~A. Athaluri, S.~Singh, {ChatGPT} in medicine: an overview of its
  applications, advantages, limitations, future prospects, and ethical
  considerations, Frontiers in artificial intelligence 6 (2023) 1169595.
\newblock \href {http://dx.doi.org/10.3389/frai.2023.1169595}
  {\path{doi:10.3389/frai.2023.1169595}}.

\bibitem{lubiana2023ten}
T.~Lubiana, R.~Lopes, P.~Medeiros, J.~C. Silva, A.~N.~A. Goncalves,
  V.~Maracaja-Coutinho, H.~I. Nakaya, Ten quick tips for harnessing the power
  of {ChatGPT} in computational biology, PLOS Computational Biology 19~(8)
  (2023) e1011319.
\newblock \href {http://dx.doi.org/10.1371/journal.pcbi.1011319}
  {\path{doi:10.1371/journal.pcbi.1011319}}.

\bibitem{choi2021chatgpt}
J.~H. Choi, K.~E. Hickman, A.~B. Monahan, D.~Schwarcz, {ChatGPT} goes to law
  school, J. Legal Educ. 71 (2021) 387.

\bibitem{hong2023chatgpt}
Z.~Hong, {ChatGPT} for computational materials science: A perspective, Energy
  Material Advances 4 (2023) 0026.
\newblock \href {http://dx.doi.org/10.34133/energymatadv.0026}
  {\path{doi:10.34133/energymatadv.0026}}.

\bibitem{deb2024chatgpt}
J.~Deb, L.~Saikia, K.~D. Dihingia, G.~N. Sastry, {ChatGPT} in the material
  design: Selected case studies to assess the potential of {ChatGPT}, Journal
  of Chemical Information and Modeling 64~(3) (2024) 799--811.
\newblock \href {http://dx.doi.org/10.1021/acs.jcim.3c01702}
  {\path{doi:10.1021/acs.jcim.3c01702}}.

\bibitem{chen2002}
L.~Q. Chen, Phase-field models for microstructure evolution, Annu. Rev. Mater.
  Res. 32 (2002) 113--140.

\bibitem{moelans2008}
N.~Moelans, B.~Blanpain, P.~Wollants, An introduction to phase-field modeling
  of microstructure evolution, Calphad 32~(2) (2008) 268 -- 294.
\newblock \href
  {http://dx.doi.org/https://doi.org/10.1016/j.calphad.2007.11.003}
  {\path{doi:https://doi.org/10.1016/j.calphad.2007.11.003}}.

\bibitem{boettinger2002}
W.~J. Boettinger, J.~A. Warren, C.~Beckermann, A.~Karma, Phase-field simulation
  of solidification, Annu. Rev. Mater. Res. 32 (2002) 163--194.
\newblock \href
  {http://dx.doi.org/https://doi.org/10.1146/annurev.matsci.32.101901.155803}
  {\path{doi:https://doi.org/10.1146/annurev.matsci.32.101901.155803}}.

\bibitem{steinbach2009}
I.~Steinbach, Phase-field models in materials science, Modelling and Simulation
  in Materials Science and Engineering 17 (2009) 073001.
\newblock \href {http://dx.doi.org/10.1088/0965-0393/17/7/073001}
  {\path{doi:10.1088/0965-0393/17/7/073001}}.

\bibitem{pf_benchmark}
A.~M. Jokisaari, P.~Voorhees, J.~E. Guyer, J.~Warren, O.~Heinonen, Benchmark
  problems for numerical implementations of phase field models, Computational
  Materials Science 126 (2017) 139--151.
\newblock \href
  {http://dx.doi.org/https://doi.org/10.1016/j.commatsci.2016.09.022}
  {\path{doi:https://doi.org/10.1016/j.commatsci.2016.09.022}}.

\bibitem{pf_benchmark2}
A.~M. Jokisaari, P.~W. Voorhees, J.~E. Guyer, J.~A. Warren, O.~G. Heinonen,
  Phase field benchmark problems for dendritic growth and linear elasticity,
  Computational Materials Science 149 (2018) 336--347.
\newblock \href
  {http://dx.doi.org/https://doi.org/10.1016/j.commatsci.2018.03.015}
  {\path{doi:https://doi.org/10.1016/j.commatsci.2018.03.015}}.

\bibitem{callister}
W.~Callister, D.~Rethwisch, Materials Science and Engineering: An Introduction,
  Wiley Plus Products Series, John Wiley \& Sons, New York, NY, 2010.

\bibitem{Cahn1958}
J.~W. Cahn, J.~E. Hilliard, Free energy of a nonuniform system. {I}.
  {I}nterfacial free energy, The Journal of chemical physics 28~(2) (1958)
  258--267.
\newblock \href {http://dx.doi.org/https://doi.org/10.1063/1.1744102}
  {\path{doi:https://doi.org/10.1063/1.1744102}}.

\bibitem{Allen1979}
S.~M. Allen, J.~W. Cahn, A microscopic theory for antiphase boundary motion and
  its application to antiphase domain coarsening, Acta Metallurgica 27~(6)
  (1979) 1085 -- 1095.
\newblock \href
  {http://dx.doi.org/https://doi.org/10.1016/0001-6160(79)90196-2}
  {\path{doi:https://doi.org/10.1016/0001-6160(79)90196-2}}.

\bibitem{kobayashi1993}
R.~Kobayashi, Modeling and numerical simulations of dendritic crystal growth,
  Physica D: Nonlinear Phenomena 63~(3-4) (1993) 410--423.
\newblock \href
  {http://dx.doi.org/https://doi.org/10.1016/0167-2789(93)90120-P}
  {\path{doi:https://doi.org/10.1016/0167-2789(93)90120-P}}.

\bibitem{hugging}
{Hugging Face}, \url{https://huggingface.co/models} (2024).

\bibitem{porter_book}
D.~A. Porter, K.~E. Easterling, Phase transformations in metals and alloys, CRC
  press, Boca Raton, FL, 2009.

\bibitem{fftw}
M.~Frigo, S.~G. Johnson, The design and implementation of {FFTW3}, Proceedings
  of the IEEE 93~(2) (2005) 216--231, special issue on ``Program Generation,
  Optimization, and Platform Adaptation''.

\bibitem{Openmp}
L.~Dagum, R.~Menon, {OpenMP: An Industry-Standard API for Shared-Memory
  Programming}, IEEE Comput. Sci. Eng. 5~(1) (1998) 46--55.
\newblock \href {http://dx.doi.org/10.1109/99.660313}
  {\path{doi:10.1109/99.660313}}.

\bibitem{Mpich}
W.~Gropp, E.~Lusk, A.~Skjellum, Using {MPI}: portable parallel programming with
  the message-passing interface, Vol.~1, MIT press, 1999.

\bibitem{Cuda}
{NVIDIA Corporation}, {NVIDIA CUDA C} programming guide, version 3.2 (2010).

\bibitem{farzadi2008}
A.~Farzadi, M.~Do-Quang, S.~Serajzadeh, A.~Kokabi, G.~Amberg, Phase-field
  simulation of weld solidification microstructure in an {Al-Cu} alloy,
  Modelling and Simulation in Materials Science and Engineering 16~(6) (2008)
  065005.
\newblock \href {http://dx.doi.org/10.1088/0965-0393/16/6/065005}
  {\path{doi:10.1088/0965-0393/16/6/065005}}.

\bibitem{ghosh2017pccp}
S.~Ghosh, A.~Mukherjee, T.~Abinandanan, S.~Bose, Particles with selective
  wetting affect spinodal decomposition microstructures, Physical Chemistry
  Chemical Physics 19~(23) (2017) 15424--15432.
\newblock \href {http://dx.doi.org/https://doi.org/10.1039/C7CP01816A}
  {\path{doi:https://doi.org/10.1039/C7CP01816A}}.

\bibitem{ghosh2020chemical}
S.~Ghosh, A.~Mukherjee, R.~Arroyave, J.~F. Douglas, Impact of particle arrays
  on phase separation composition patterns, The Journal of chemical physics
  152~(22) (2020) 224902.
\newblock \href {http://dx.doi.org/https://doi.org/10.1063/5.0007859}
  {\path{doi:https://doi.org/10.1063/5.0007859}}.

\bibitem{ghosh2024_fractal}
S.~Ghosh, J.~F. Douglas, {Phase separation in the presence of fractal
  aggregates}, The Journal of Chemical Physics 160~(10) (2024) 104903.
\newblock \href {http://dx.doi.org/10.1063/5.0190196}
  {\path{doi:10.1063/5.0190196}}.

\bibitem{ghosh2022tusas}
S.~Ghosh, C.~K. Newman, M.~M. Francois, Tusas: A fully implicit parallel
  approach for coupled phase-field equations, Journal of Computational Physics
  448 (2022) 110734.
\newblock \href {http://dx.doi.org/https://doi.org/10.1016/j.jcp.2021.110734}
  {\path{doi:https://doi.org/10.1016/j.jcp.2021.110734}}.

\bibitem{steinbach2013phase}
I.~Steinbach, Phase-field model for microstructure evolution at the mesoscopic
  scale, Annual Review of Materials Research 43 (2013) 89--107.
\newblock \href
  {http://dx.doi.org/https://doi.org/10.1146/annurev-matsci-071312-121703}
  {\path{doi:https://doi.org/10.1146/annurev-matsci-071312-121703}}.

\bibitem{voorhees1985}
P.~W. Voorhees, The theory of {O}stwald ripening, J. Stat. Phys 38~(1-2) (1985)
  231--252.
\newblock \href {http://dx.doi.org/https://doi.org/10.1007/BF01017860]}
  {\path{doi:https://doi.org/10.1007/BF01017860]}}.

\bibitem{lee2013comparison}
S.~Lee, C.~Lee, H.~G. Lee, J.~Kim, Comparison of different numerical schemes
  for the cahn-hilliard equation, Journal of the Korean Society for Industrial
  and Applied Mathematics 17~(3) (2013) 197--207.
\newblock \href {http://dx.doi.org/https://doi.org/10.12941/jksiam.2013.17.197}
  {\path{doi:https://doi.org/10.12941/jksiam.2013.17.197}}.

\bibitem{provatasbook}
N.~Provatas, K.~Elder, Phase-field methods in materials science and
  engineering, John Wiley \& Sons, New York, NY, 2011.

\bibitem{ghosh20183d}
S.~Ghosh, N.~Ofori-Opoku, J.~E. Guyer, Simulation and analysis of $\gamma$-ni
  cellular growth during laser powder deposition of ni-based superalloys,
  Computational Materials Science 144 (2018) 256--264.
\newblock \href
  {http://dx.doi.org/https://doi.org/10.1016/j.commatsci.2017.12.037}
  {\path{doi:https://doi.org/10.1016/j.commatsci.2017.12.037}}.

\bibitem{ghosh2017_eutectic}
S.~Ghosh, M.~Plapp, Influence of interphase boundary anisotropy on bulk
  eutectic solidification microstructures, Acta Materialia 140 (2017) 140--148.
\newblock \href
  {http://dx.doi.org/https://doi.org/10.1016/j.actamat.2017.08.023}
  {\path{doi:https://doi.org/10.1016/j.actamat.2017.08.023}}.

\bibitem{hillert1965theory}
M.~Hillert, On the theory of normal and abnormal grain growth, Acta
  metallurgica 13~(3) (1965) 227--238.
\newblock \href
  {http://dx.doi.org/https://doi.org/10.1016/0001-6160(65)90200-2}
  {\path{doi:https://doi.org/10.1016/0001-6160(65)90200-2}}.

\bibitem{fan1997computer}
D.~Fan, L.-Q. Chen, Computer simulation of grain growth using a continuum field
  model, Acta Materialia 45~(2) (1997) 611--622.
\newblock \href
  {http://dx.doi.org/https://doi.org/10.1016/S1359-6454(96)00200-5}
  {\path{doi:https://doi.org/10.1016/S1359-6454(96)00200-5}}.

\bibitem{kurz2019progress}
W.~Kurz, D.~J. Fisher, R.~Trivedi, Progress in modelling solidification
  microstructures in metals and alloys: dendrites and cells from 1700 to 2000,
  International Materials Reviews 64~(6) (2019) 311--354.
\newblock \href
  {http://dx.doi.org/https://doi.org/10.1080/09506608.2018.1537090}
  {\path{doi:https://doi.org/10.1080/09506608.2018.1537090}}.

\bibitem{kurz2021progress}
W.~Kurz, M.~Rappaz, R.~Trivedi, Progress in modelling solidification
  microstructures in metals and alloys. part ii: dendrites from 2001 to 2018,
  International Materials Reviews 66~(1) (2021) 30--76.
\newblock \href
  {http://dx.doi.org/https://doi.org/10.1080/09506608.2020.1757894}
  {\path{doi:https://doi.org/10.1080/09506608.2020.1757894}}.

\bibitem{Echebarria2004}
B.~Echebarria, R.~Folch, A.~Karma, M.~Plapp, Quantitative phase-field model of
  alloy solidification, Physical {R}eview {E} 70~(6) (2004) 061604.
\newblock \href {http://dx.doi.org/https://doi.org/10.1103/PhysRevE.70.061604}
  {\path{doi:https://doi.org/10.1103/PhysRevE.70.061604}}.

\bibitem{Trevor2017}
T.~Keller, G.~Lindwall, S.~Ghosh, L.~Ma, B.~Lane, F.~Zhang, U.~R. Kattner,
  E.~A. Lass, J.~C. Heigel, Y.~Idell, M.~E. Williams, A.~J. Allen, J.~E. Guyer,
  L.~E. Levine, Application of {F}inite {E}lement, {P}hase-field, and
  {CALPHAD}-based {M}ethods to {A}dditive {M}anufacturing of {N}i-based
  {S}uperalloys, Acta {M}aterialia 139 (2017) 244--253.
\newblock \href
  {http://dx.doi.org/https://doi.org/10.1016/j.actamat.2017.05.003}
  {\path{doi:https://doi.org/10.1016/j.actamat.2017.05.003}}.

\bibitem{Matlab}
MATLAB, Version R2018a, The MathWorks Inc., Natick, Massachusetts, 2018.

\bibitem{gpt4_turbo}
{GPT-4 Turbo}, \url{https://platform.openai.com/docs/models/overview/} (29
  March 2024).

\bibitem{gpt4o}
{GPT-4o}, \url{https://openai.com/index/hello-gpt-4o/} (May 13, 2024).

\bibitem{tweet}
{OpenAI@OpenAI}, \url{https://twitter.com/OpenAI/status/1773738074041717109}
  (29 March 2024).

\end{thebibliography}

\end{document}